\documentclass[secnumarabic,aps,pra,amsfonts,twocolumn,%
showkeys,floatfix%
]{revtex4-1}

\usepackage{bm,amsfonts,amsthm}
\usepackage{sidecap}
\usepackage{placeins}
\usepackage{graphicx}
\newcommand{\gl}[1]{(\ref{#1})}

\newtheorem{thm}{Theorem}

\renewcommand{\thetable}{\arabic{table}}

\begin{document}
\title{Structural distortion in
  antiferromagnetic BaFe$_2$As$_2$ as a result of time-inversion symmetry} 
\author{Ekkehard Kr\"uger}
\author{Horst P. Strunk} \affiliation{Institut f\"ur
  Materialwissenschaft, Materialphysik, Universit\"at Stuttgart,
  D-70569 Stuttgart, Germany}
%
\date{\today}
\begin{abstract}
  As reported by Q. Huang et al. [Phys. Rev. Lett. 101, 257003
  (2008)], neutron diffraction studies show an onset of
  antiferromagnetic order in BaFe$_2$As$_2$ associated with a
  tetragonal-to-orthorhombic distortion. We determine the group $Cmca$
  as the space group of antiferromagnetic BaFe$_2$As$_2$ and identify
  a roughly half-filled energy band of BaFe$_2$As$_2$ with Bloch
  functions of special symmetry as magnetic band. As explained by the
  group-theoretical nonadiabatic Heisenberg model, the electrons in
  this narrow band may lower their Coulomb correlation energy by
  producing just the experimentally observed antiferromagnetic state
  if this state does not violate group-theoretical
  principles. However, in {\em undistorted} BaFe$_2$As$_2$ the
  time-inversion symmetry of the system interferes with the stability
  of the antiferromagnetic state. Nevertheless, it can be stabilized
  by a structural distortion of BaFe$_2$As$_2$ going beyond the
  magnetostriction. We derive two possible structural distortions
  stabilizing the antiferromagnetic state. These distortions are
  described by their space groups and consist in mere displacements of
  the Fe atoms.
\end{abstract}

\maketitle

\section{Introduction}

Neutron diffraction investigations of BaFe$_2$As$_2$~\cite{huang2}
revealed a tetragonal to orthorhombic distortion associated with an
antiferromagnetic order where the magnetic and the structural
transitions occur at the same temperature. The present paper analyses
this obvious coincidence of magnetic and structural phase transitions
in view of a group-theoretical approach which has already shed insight
into the reasons for both magnetic and superconducting phases. This
approach bases on the observation that the occurrence of a magnetic or
superconducting state is always related to the existence of a narrow,
almost half-filled magnetic or superconducting band of well-defined
symmetry in the band structure of the respective material.  Examples
are the magnetic materials Cr~\cite{ea}, Fe~\cite{ef},
La$_2$CuO$_4$~\cite{josla2cuo4}, YBa$_2$Cu$_3$O$_6$~\cite{ybacuo6},
and LaFeAsO~\cite{lafeaso1}, the (high-T$_c$) superconductors
La$_2$CuO$_4$~\cite{josla2cuo4}, YBa$_2$Cu$_3$O$_7$,
MgB$_2$~\cite{josybacuo7}, and doped LaFeAsO~\cite{lafeaso2}, and
numerous elemental superconductors~\cite{es2,josm}.

The notion of the used approach, the so-called nonadiabatic Heisenberg
model, is that the electronic system in a narrow, roughly half-filled
magnetic or superconducting band can lower its Coulomb correlation
energy by condensing into an atomic-like state. However, for a
consistent description of this atomic-like state we must also take
into account the motion of the atomic cores following the electronic
motion. Fortunately, we may assume that the very complex nonadiabatic
localized functions representing the related localized states have the
same symmetry and spin-dependence as the best localized,
symmetry-adapted and, in the case of a superconducting band,
spin-dependent Wannier functions of the magnetic or superconducting
band. As a consequence of the special symmetry and spin-dependence of
the localized states of a magnetic or superconducting band, the system
is necessarily magnetic or a superconductor, respectively, in the
atomic-like state~\cite{enhm}.  In case of magnetic bands, however, an
irrefutable condition is that related magnetic state does not conflict
with the time-inversion symmetry of the system.

In Sec.~\ref{sec:64} we shall identify the group $Cmca$ (64) as the
space group of the antiferromagnetic structure observed in
BaFe$_2$As$_2$ (the number in parenthesis is the international number)
and shall determine the magnetic group of this structure. In the
following Sec.~\ref{sec:magneticband} we will verify the existence of
a magnetic band related to the observed antiferromagnetic structure in
the band structure of BaFe$_2$As$_2$.

In Sec.~\ref{sec:stability} we shall show that this antiferromagnetic
state is in fact unstable in {\em undistorted} BaFe$_2$As$_2$, because
it conflicts with the time-inversion symmetry of the system. This statement 
means that the antiferromagnetic state exists in
BaFe$_2$As$_2$ only, if it is accompanied by a distortion of the crystal
going beyond the magnetostriction. Such distortions stabilizing an
antiferromagnetic state have already been detected in
La$_2$CuO$_4$~\cite{josla2cuo4} and LaFeAsO~\cite{lafeaso1}. In
Sec.~\ref{sec:subgroups} we will determine the space groups of those
distortions for BaFe$_2$As$_2$ that stabilize the observed
antiferromagnetic structure.

\begin{figure*}
\begin{minipage}[c]{.45\textwidth}
\includegraphics[width=.9\textwidth,angle=0]{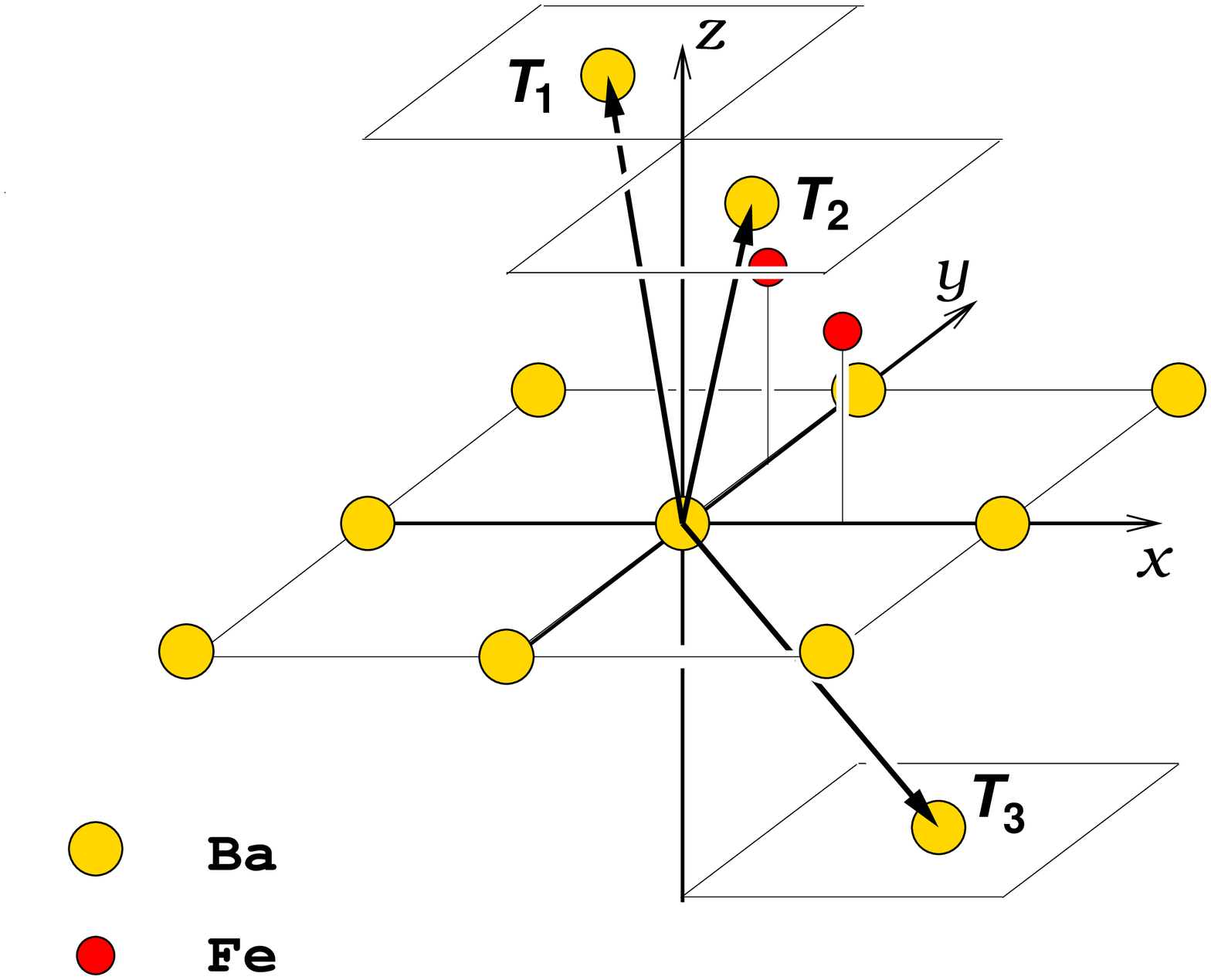}%
\begin{center}
(a)\ $I4/mmm = \Gamma^v_qD^{17}_{4h}$ (139)
\end{center}
\end{minipage}
\begin{minipage}{.45\textwidth}
\includegraphics[width=.9\textwidth,angle=0]{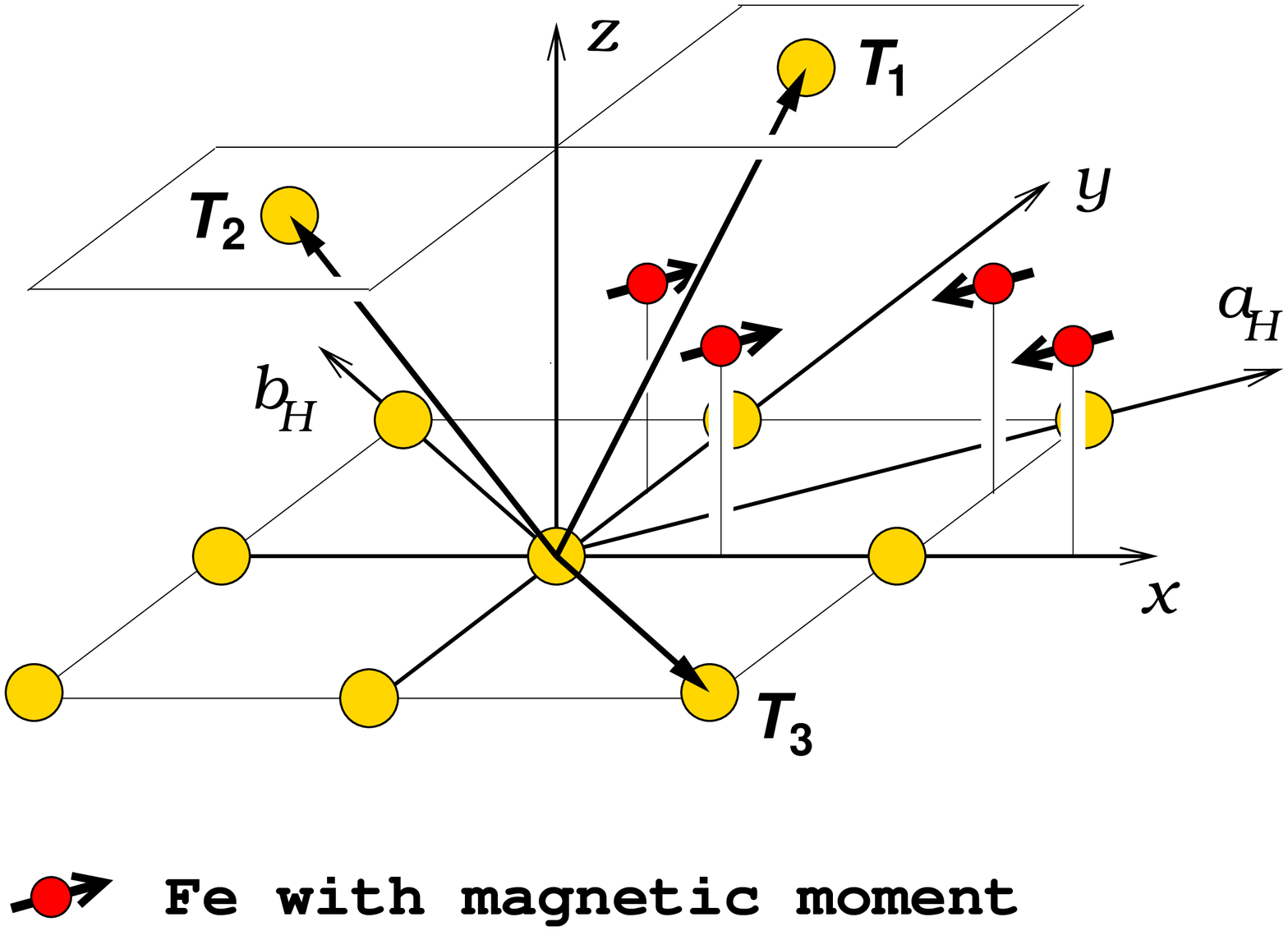}%
\begin{center}
\vspace{.5cm}
(b)\ $Cmca = \Gamma^b_oD^{18}_{2h}$ (64)
\end{center}
\end{minipage}
\vspace{1cm}

\begin{minipage}[b]{.45\textwidth}
\includegraphics[width=.9\textwidth,angle=0]{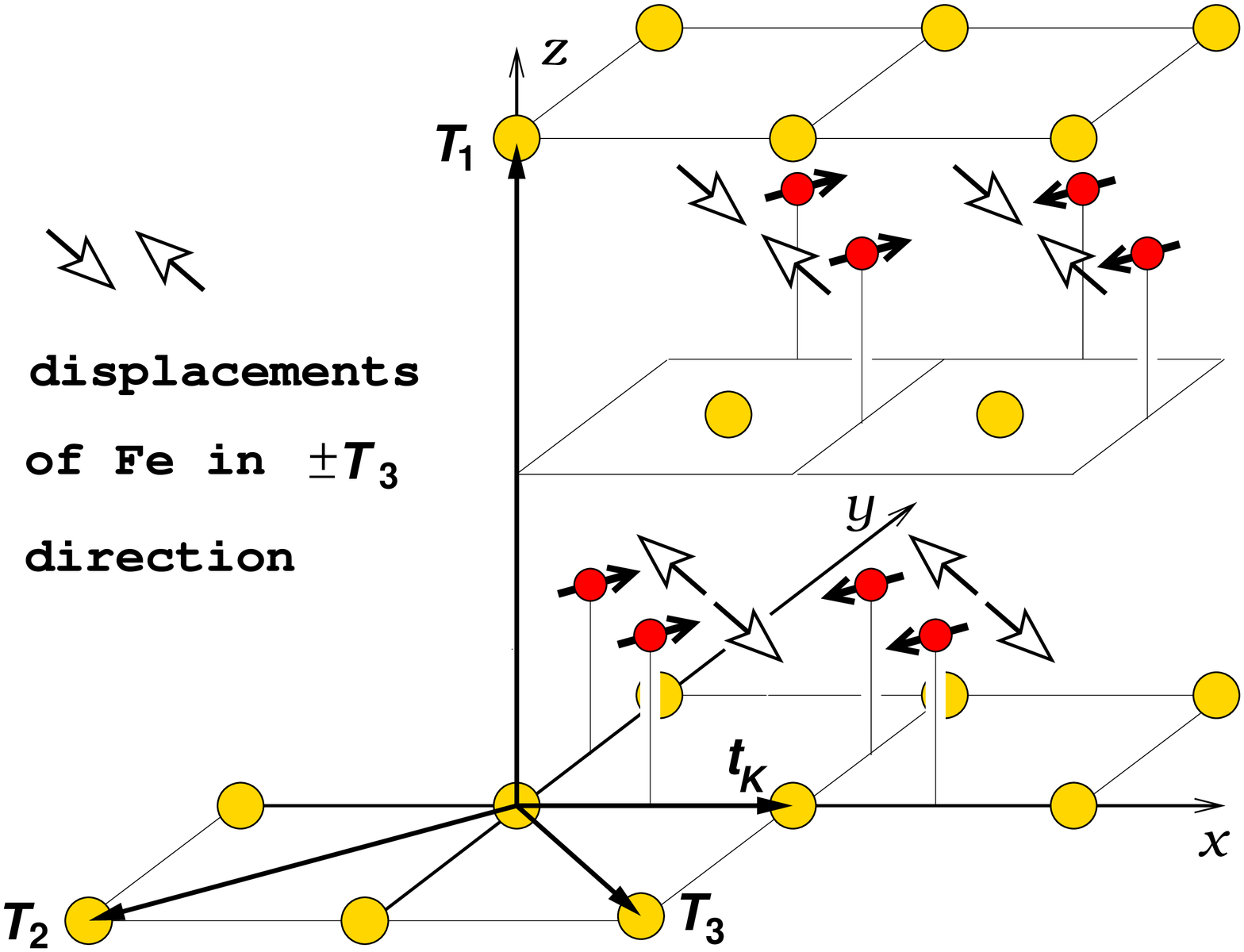}%
\begin{center}
(c) $Pnma = \Gamma_oD^{16}_{2h}$ (62)
\vspace{-3cm}
\end{center}
\end{minipage}
\begin{minipage}{.45\textwidth}
\includegraphics[width=.9\textwidth,angle=0]{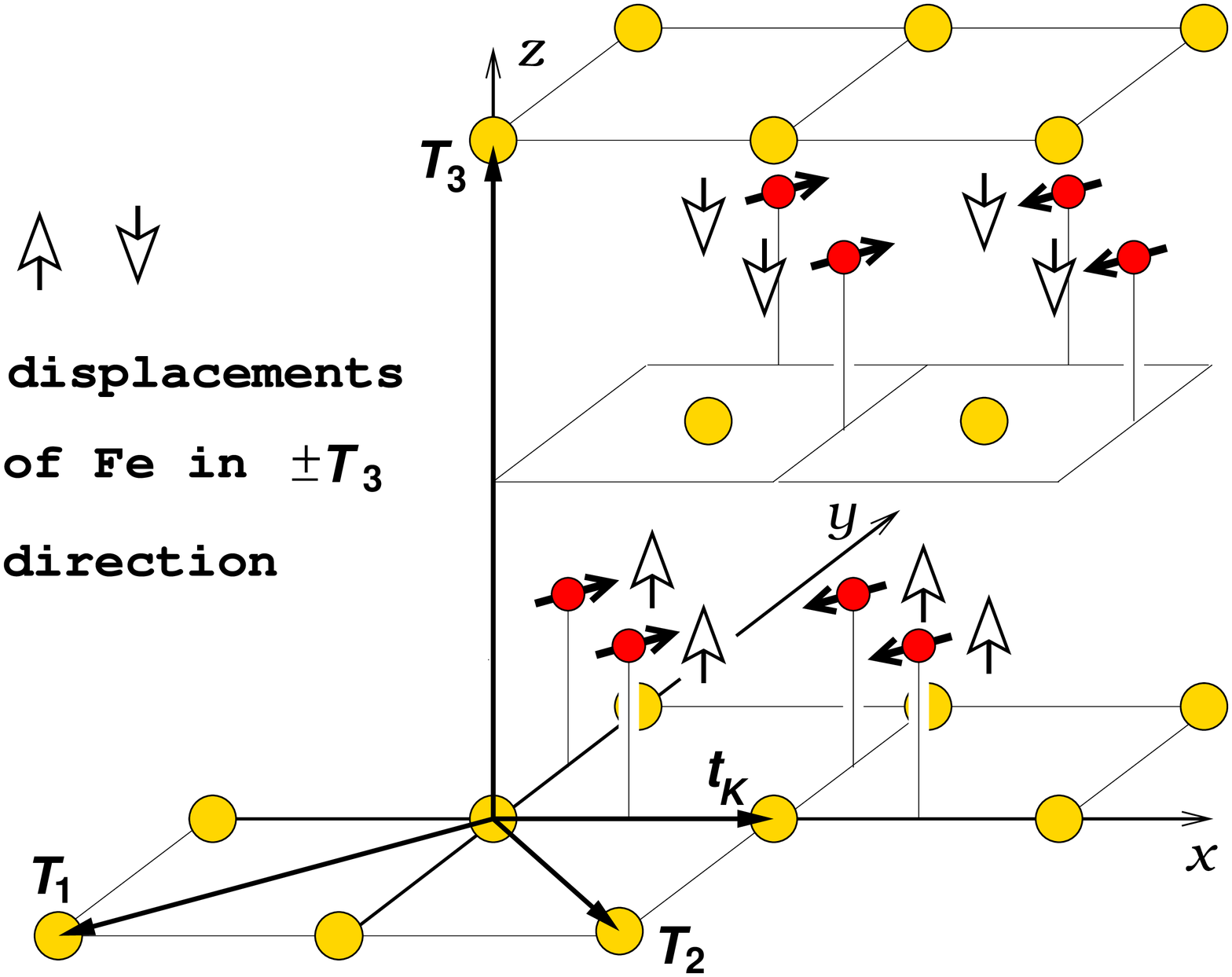}%
\begin{center}
(d) $Pbam = \Gamma_oD^{9}_{2h}$ (55)
\end{center}
\end{minipage}
\caption{Paramagnetic (a), undistorted antiferromagnetic (b)
  BaFe$_2$As$_2$, and structural distortions (c) and (d) stabilizing
  the antiferromagnetic structure.  Great (yellow) circles represent
  the Ba, small (red) circles the Fe atoms.  For reasons of clarity
  the As atoms are not shown. While sufficient Ba atoms are depicted
  to recognize the orientation of the crystal, the Fe atoms are shown
  only within one unit cell.  In order to facilitate a comparison of
  the instructive Fig.~3 of Ref.~\cite{huang2} with our figures, we
  have denoted in Fig.~(b) by $a_H$ and $b_H$ the directions of the
  $a$ and $b$ axis used in Ref.~\cite{huang2}.  The coordinate systems
  define the symmetry operations $\{R|pqr\}$ as used in this
  paper. They are written in the Seitz notation detailed in the
  textbook of Bradley and Cracknell \cite{bc}: $R$ stands for a point
  group operation and $pqr$ denotes the subsequent translation $\bm
  t$.  The point group operation is related to the $x, y, z$
  coordinate system (as described in Fig. 2 of Ref.~\cite{lafeaso1}),
  and $pqr$ stands for the translation $\bm t = p\bm T_1 + q\bm T_2 +
  r\bm T_3$. While the orientation of the crystal and the $x, y, z$
  coordinate system are fixed, the basic translations $\bm T_i$ are
  different in each structure. The distortions (c) and (d) may be
  realized by the depicted (slight) displacements of the Fe atoms in
  exact $\pm \bm T_3$ direction in each case [where the basic 
  translations $\bm T_3$ are different in (c) and (d)]. It
  should be noted that the given displacements of the Fe atoms may be
  accompanied by additional displacements of the As or Ba atoms
  consistent with the respective magnetic group. The translation $\bm
  t_K$ denotes the translation associated with the time inversion
  $K$. $\bm t_K$ is a translation of the paramagnetic crystal that
  connects two Fe atoms with antiparallel magnetic moments and the
  {\em same} displacement.
\label{fig:structures}
}
\end{figure*}

We shall find the two possible distortions (described by their space
groups) depicted in Figs.~\ref{fig:structures} (c) and (d). Both
distortions can be realized by a mere displacement of the iron atoms.
We tend to the distortion in Fig.~\ref{fig:structures} (c) to be
most likely realized in BaFe$_2$As$_2$ because it clearly confirms the
experimental observation by Huang et al.~\cite{huang2} that the $a$
and $b$ axes (denoted here in Fig.~\ref{fig:structures} (b) by $a_H$ and
$b_H$, respectively) have different lengths in the antiferromagnetic
phase.  Furthermore, this distortion is realized by nearest-neighbor Fe
atoms that are oppositely displaced. Thus, the displacements are
effected by a nearest-neighbor interaction between the Fe atoms which
may be more efficient than an interaction between the Fe atoms in
different layers as it appears to be effective in
Fig.~\ref{fig:structures} (d).

The nonadiabatic Heisenberg model does not distinguished between
orbital and spin moments.  Therefore, we always speak of ``magnetic
moments'' which may consist of both orbital and spin moments.

\section{The magnetic group of the experimentally observed magnetic
  structure}
\label{sec:64}
Fig.~\ref{fig:structures} (a) shows the two Fe atoms in the
paramagnetic and Fig.~\ref{fig:structures} (b) the four Fe atoms in
the antiferromagnetic unit cell of BaFe$_2$As$_2$. By inspection we
may recognize that Fig.~\ref{fig:structures} (b) displays the
experimentally determined~\cite{huang2} antiferromagnetic structure
because by application of the given basic translations $\bm T_1, \bm
T_2, \bm T_3$ to the four Fe atoms in the unit cell we obtain just
the magnetic structure presented in Fig.~3 of Ref.~\cite{huang2}.

Removing from the space group $I4/mmm$ of BaFe$_2$As$_2$ all the
symmetry operations not leaving invariant the magnetic moments of the
Fe atoms [as depicted in Fig.~\ref{fig:structures} (b)], we obtain the
group $Cmca = \Gamma^b_oD^{18}_{2h}$ (64) as the space group of the
antiferromagnetic structure in undistorted BaFe$_2$As$_2$. This
may be proved using Figs.~\ref{fig:structures} (a) and (b) and also
the ``generating elements''
\begin{equation} 
\label{eq:1}
\{C_{2b}|\textstyle\frac{1}{2}\textstyle\frac{1}{2}\textstyle\frac{1}{2}\}, 
\{C_{2a}|\textstyle\frac{1}{2}\textstyle\frac{1}{2}\textstyle\frac{1}{2}\},
\text{ and } \{I|000\}
\end{equation} 
of $Cmca$, as it was detailed in Sec.~3.1 of Ref~\cite{lafeaso1}. The
generating elements are taken from Table 3.7 of Ref.~\cite{bc},
however, they are written in this paper in the coordinate
system defined by Fig.~\ref{fig:structures} (b), cf. the notes to
Table~\ref{tab:rep64}.

Furthermore, by inspection of Fig.~\ref{fig:structures} (b) we see
that the antiferromagnetic structure in undistorted BaFe$_2$As$_2$ is
invariant under the anti-unitary operation
$\{K|\frac{1}{2}\overline{\frac{1}{2}}\frac{1}{2}\}$, where $K$
denotes the operator of time inversion reversing all the magnetic
moments and leaving invariant the positions of the atoms.  Thus, the
magnetic group $M_{64}$ of the experimentally determined magnetic
structure in undistorted BaFe$_2$As$_2$ may be written as
\begin{equation}
  \label{eq:3}
  M_{64} = Cmca + \{K|\textstyle\frac{1}{2}\overline{\frac{1}{2}}\frac{1}{2}\}Cmca.
\end{equation}

 \begin{figure*}[t]
 \includegraphics[width=.65\textwidth,angle=-90]{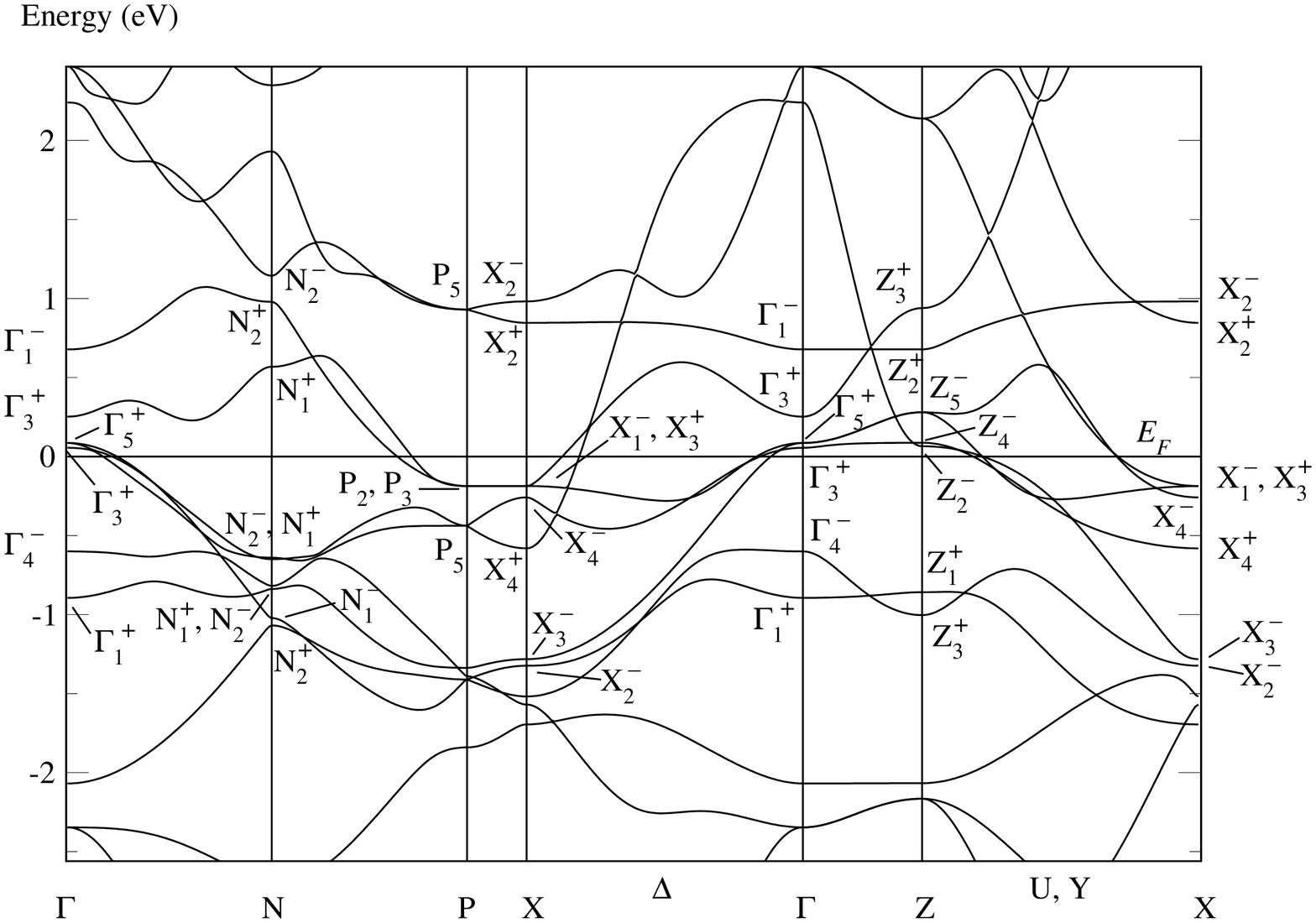}
 \caption{Band structure of tetragonal BaFe$_2$As$_2$ as calculated by the 
   FHI-aims program \cite{blum1,blum2}, using the structure parameters given 
   in Ref.~\cite{huang2}, with symmetry labels determined by
   the authors. The notations of the points and lines of symmetry in the 
   Brillouin zone for $\Gamma^v_q$ follow Fig. 3.10. (b) of Ref.~\cite{bc},
   and the symmetry labels are defined by Table~\ref{tab:rep139}.  
   \label{fig:bandstr1}
 }
 \end{figure*}


\section{Magnetic band of $\text{BaFe$_2$As$_2$}$}
\label{sec:magneticband}

In this section we show that in the band structure of paramagnetic
BaFe$_2$As$_2$ we find a magnetic band~\cite{enhm} related to the
magnetic group $M_{64}$ determined in the preceding section. 

The band structure of paramagnetic BaFe$_2$As$_2$ (with the space
group $I4/mmm$) is depicted in Fig.  ~\ref{fig:bandstr1}.  There are
Bloch functions near the Fermi level characterized by the
representations
\begin{equation}
  \label{eq:2}
\Gamma^+_5, \Gamma^+_3; Z^-_5, Z^-_4; X^-_1, X^+_3, X^-_4, X^+_4; N^+_1, N^-_2.
\end{equation}
Folding the band structure into the Brillouin zone of the space group $Cmca$ of
the antiferromagnetic structure in the undistorted crystal [depicted in
Fig.~\ref{fig:structures} (b)], the representations~\gl{eq:2} of the Bloch
functions transform as
\begin{equation}
  \label{eq:7}
\begin{array}{lclp{.3cm}lcl}
\Gamma^+_5 & \rightarrow & \Gamma^+_2 + \underline{\Gamma^+_3} && 
\Gamma^+_3 & \rightarrow & \underline{\Gamma^+_4}\\[.2cm] 
Z^-_5 & \rightarrow & Y^-_2 + \underline{Y^-_3}&&  
Z^-_4 & \rightarrow & \underline{Y^-_4}\\[.2cm]  
X^-_1 & \rightarrow & \underline{\Gamma^-_4}&&  
X^+_3 & \rightarrow & \underline{Y^+_4}\\[.2cm]  
X^-_4 & \rightarrow & \underline{\Gamma^-_3}&&  
X^+_4 & \rightarrow & \underline{Y^+_3}\\[.2cm]  
N^+_1 & \rightarrow & \underline{R^-_1} + \underline{R^-_2} &&  
N^-_2 & \rightarrow & \underline{R^+_1} + \underline{R^+_2} \\[.2cm]  
\end{array}
\end{equation}
see Table~\ref{tab:falten139_64}. The underlined representations form
a magnetic band listed in Table~\ref{tab:wf64}, namely band 2. 
The representations $Z_1 + Z_2$ as well as $T_1 + T_2$ are absent
in~\gl{eq:7} though they belong to the magnetic band, too. They may be
determined by the compatibility relations given in
Table~\ref{tab:komprel139} and by Table~\ref{tab:falten139_64}.

\begin{figure*}[!]
 \includegraphics[width=.55\textwidth,angle=-90]{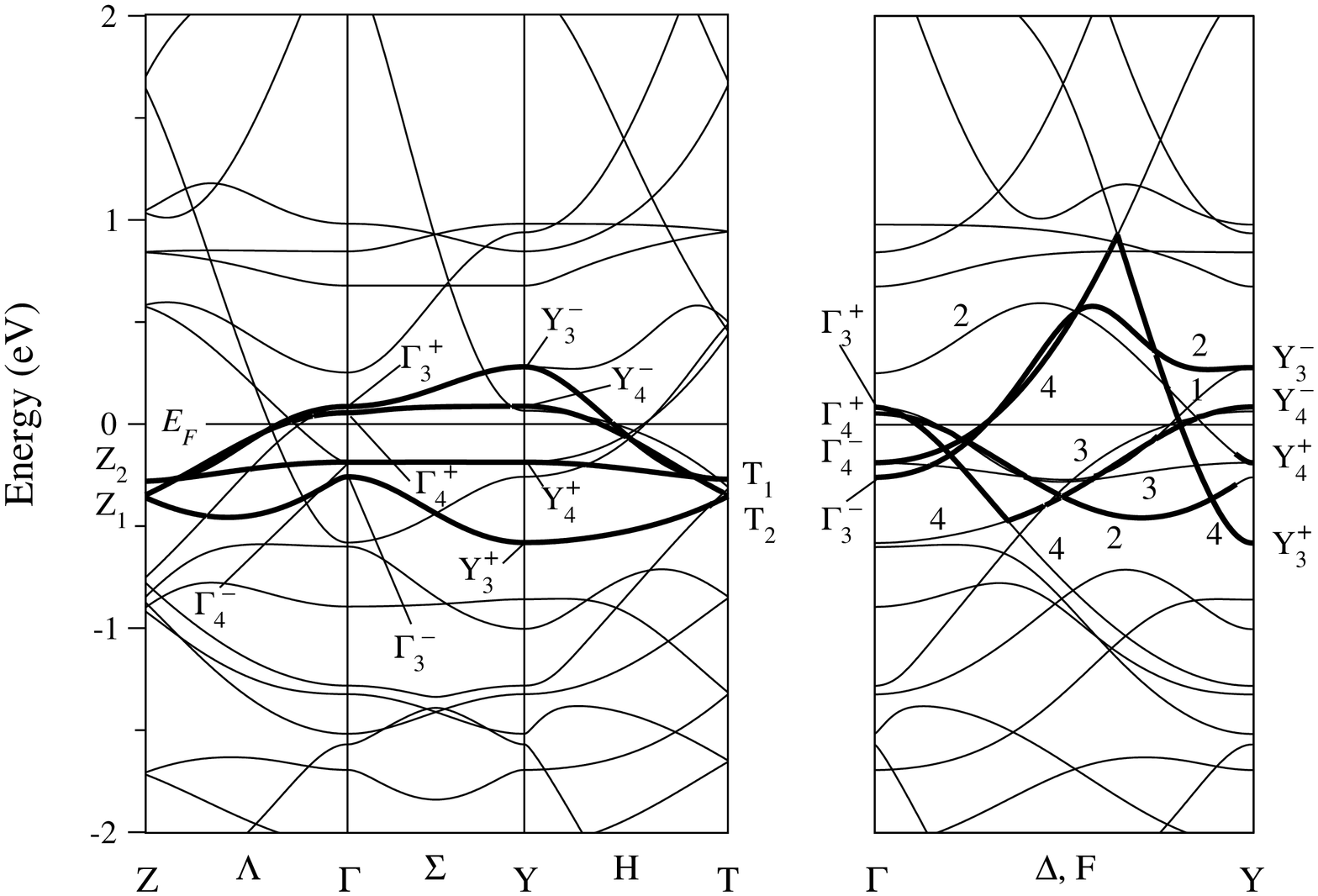}%
 \caption{The band structure of tetragonal BaFe$_2$As$_2$ (as given in 
   Fig.~\ref{fig:bandstr1}) folded into the Brillouin zone for the
Bravais lattice $\Gamma^b_o$ of the space group $Cmca$ of the 
antiferromagnetic structure. The bold line 
shows the magnetic band consisting of four branches with symmetry labels 
that can be identified from Table~\ref{tab:rep64}.
The notations of the points and lines of symmetry follow Fig. 3.6. (a) of 
Ref.~\cite{bc}. The numbers labeling the branches at the line $\Delta$, F 
characterize 
the related representations; for instance, the number 2 stands for $\Delta_2$ and 
F$_2$ (as defined in Table~\ref{tab:komprel64}). In the surroundings of the sharp
peaks at the line $\Delta$, F (between $\Gamma$ and $Y
(\overline{\frac{1}{2}}\frac{1}{2}0)$), 
the Wannier functions are build from Bloch-like functions (equation (1.4) of
Ref.~\cite{enhm}) of $\Delta_4$, F$_4$ symmetry that are linear combinations of two $\Delta_4$,
F$_4$ Bloch functions arising from different lines in the Brillouin zone for the paramagnetic 
lattice $\Gamma^v_q$. These linear combinations 
smooth away the peaks. Alike, the
little jump between two $\Delta_2$ (F$_2$) branches near Y$^+_4$ is bridged 
by a linear combination of the related $\Delta_2$ (F$_2$) Bloch
functions. Thus, this little jump is physically meaningless since it does not cross 
the Fermi level. 
  \label{fig:bandstr2}
}
 \end{figure*}

The run of the magnetic band may not be visualized until the
paramagnetic band structure is folded into the Brillouin zone of the
antiferromagnetic structure lying diagonally within the Brillouin zone
for the paramagnetic phase. Always two lines in the paramagnetic
Brillouin zone are equivalent to one line in the antiferromagnetic
Brillouin zone, see Table~\ref{tab:falten}.  The folded band structure
with the magnetic band (highlighted by the bold line) is depicted in
Fig.~\ref{fig:bandstr2}.

The Bloch functions of the magnetic band can be unitarily transformed
into Wannier functions that are
\begin{itemize}
\item as well localized as possible, 
\item centered at the Fe atoms,
\item and symmetry-adapted to the magnetic group $M_{64}$ in Eq.~\gl{eq:3}, 
\end{itemize}
see the notes to Table~\ref{tab:wf64}. 

The magnetic band in BaFe$_2$As$_2$ is roughly half-filled and
extremely narrow as compared with the other bands in the band
structure of this material. Thus, this band provides optimal features
enabling the electrons to lower their Coulomb correlation
energy~\cite{enhm,ea,ef,lafeaso1} by activating an exchange mechanism
producing the experimentally determined antiferromagnetic
structure. However, we shall show in the following
Sec.~\ref{sec:stability} that in {\em undistorted} BaFe$_2$As$_2$ this
mechanism is blocked by the time-inversion symmetry of the system.

\section{Stability of the antiferromagnetic structure}
\label{sec:stability}
In this section we show that the antiferromagnetic state with the
space group $Cmca$ requires a (slight) distortion of BaFe$_2$As$_2$ to
form a different space group with restored time-inversion symmetry.

\subsection{The antiferromagnetic structure is not stable in
  undistorted BaFe$_2$As$_2$}
\label{sec:conditions}
Assume that the magnetic state $|G\rangle$ is an eigenstate of a
Hamiltonian $\widetilde H$ commuting with the operator $K$ of time
inversion,
\begin{equation}
  \label{eq:18}
 [\widetilde H, K] = 0,
\end{equation}
as it is the case within the nonadiabatic Heisenberg model, see
Sec.~III.C. of Ref.~\cite{ea}.  Nevertheless, we get a state different
from $|G\rangle$ when we apply $K$ to $|G\rangle$,
\begin{equation}
  \label{eq:17}
  K|G\rangle \neq |G\rangle,
\end{equation}
since $K$ reverses the magnetic moments.  Consequently~\cite{ea},
$|G\rangle$ and $K|G\rangle$ are basis functions of a two-dimensional
{\em irreducible} corepresentations $\widetilde D$ of the magnetic
group $\widetilde M$ of $\widetilde H$. $\widetilde M$ may be written
as
\begin{equation}
  \label{eq:16}
\widetilde M = M + KM,  
\end{equation}
where $M$ denotes the magnetic group 
\begin{equation}
  \label{eq:15}
M = S + \{K|\bm t_K\}S 
\end{equation}
of the magnetic state and $S$ is the related space group. 

However, when we restrict the symmetry operations to the
subgroup $M$ of $\widetilde M$, then $\widetilde D$ must subduce two {\em
  one-dimensional} corepresentation $D$ of $M$, because, e.g., we have 
\begin{equation}
  \label{eq:19}
\{K|\bm t_K\}|G\rangle = c\cdot|G\rangle,
\end{equation}
since the anti-unitary operator $\{K|\bm t_K\}$ belongs to the magnetic
group $M$ of the magnetic state. $c$ is a complex number with 
$|c| = 1$.

If the magnetic group $\widetilde M$ does not possess such suitable
corepresentations $\widetilde D$, the related magnetic state
is necessarily unstable.  Fortunately, we may recognize {\em by the
  known representations of the space group $S$ alone} whether or not
$\widetilde M$ possesses corepresentations $\widetilde D$ allowing a
stable magnetic state. This may be carried out by means of
Theorem 4.1 in Ref.~\cite{lafeaso1} reading as follows:

\begin{thm}
\label{theorem}
The magnetic group $\widetilde M$~\gl{eq:16} possesses
corepresentations allowing a stable magnetic state with the space group
$S$ if $S$ has at least one one-dimensional single-valued
representation
\begin{enumerate} 
\item 
following case (a) with respect to the magnetic group 
$S + \{K|\bm t_K\}S$~\gl{eq:15} of the magnetic state and
\item 
following case (c) with
respect to the magnetic group $S + KS$.
\end{enumerate}
\end{thm}
The cases (a) and (c) are defined by Eqs.\ (7.3.45) and (7.3.47),
respectively, of Ref.~\cite{bc}, and are given for the representations
relevant in this paper in Tables~\ref{tab:rep64},~\ref{tab:rep62},
and~\ref{tab:rep55}. This theorem led already to an understanding of
the symmetry of the incommensurate spin-density-wave states in
Cr~\cite{ea}.

At first sight, the representations at point $R$ in the Brillouin for
$Cmca$ appear to comply with both conditions (i) and (ii), see
Table~\ref{tab:rep64}. However, not all the space group operations of
$Cmca$ are comprised within the little group at $R$, and,
consequently, the magnetic group $\widetilde M$~\gl{eq:16} related to
$M = M_{64}$~\gl{eq:3} does not possess corepresentations allowing a
stable antiferromagnetic state with the space group $Cmca$.

\subsection{Reaching stability of the antiferromagnetic state by
slight distortion of the BaFe$_2$As$_2$ structure}
\label{sec:subgroups}
The experimentally observed antiferromagnetic structure may be
stabilized by a (small) distortion of the crystal that turns the space
group $Cmca$ into an ``allowed'' subgroup of $Cmca$ possessing
representations that allow the formation of a stable antiferromagnetic
structure. In this context we considered all the space groups
\begin{enumerate}
\item 
  leaving invariant the experimentally observed magnetic structure
  depicted in Fig.~\ref{fig:structures} (b) and
\item 
  meeting the two conditions in Theorem~\ref{theorem}.
\end{enumerate}

In this way we found numerous allowed space groups in antiferromagnetic
BaFe$_2$As$_2$, among them also groups containing very few symmetry
operations. In the following we only consider the space groups $Pnma$
(62) and $Pbam$ (55) which are the allowed space groups of highest
symmetry in antiferromagnetic BaFe$_2$As$_2$. Both groups have the
orthorhombic-primitive Bravais lattice $\Gamma_o$ and comprise the
full orthorhombic point group $D_{2h}$.

The space groups $Pnma$ and $Pbam$ can be defined by the generating elements
\begin{equation}
  \label{eq:9}
\{C_{2a}|0\textstyle\frac{1}{2}\textstyle\frac{1}{2}\},\ 
\{C_{2z}|\textstyle\frac{1}{2}\textstyle\frac{1}{2}0\},\   
\{I|\textstyle\frac{1}{2}\textstyle\frac{1}{2}0\}, 
\end{equation}
and 
\begin{equation}
  \label{eq:11}
\{C_{2a}|\textstyle\frac{1}{2}\frac{1}{2}0\},\ 
\{C_{2z}|000\},\ 
\{I|000\}, 
\end{equation}
respectively, as given in Table 3.7 of Ref.~\cite{bc}. Here, however,
they are written in the coordinate systems given by
Figs.~\ref{fig:structures} (c) and (d), respectively.

The symmetry operations of both $Pnma$ and $Pbam$ leave invariant the
atoms of BaFe$_2$As$_2$ and the antiferromagnetic structure depicted
in Fig.~\ref{fig:structures} (b).  This may be proved starting from
the generating elements of $Pnma$ and $Pbam$, as it was detailed in
Sec.~5.1 of Ref~\cite{lafeaso1}.

In contrast to the group $Cmca$, both groups $Pnma$ and $Pbam$ possess
one-dimensional representations satisfying both conditions (i) and
(ii) of Theorem~\ref{theorem}. These are the representations at point
$U$ in the Brillouin zone for $Pnma$ and at points $S$ and $R$ in the
Brillouin zone for $Pbam$, see Tables~\ref{tab:rep62}
and~\ref{tab:rep55}, respectively. In any case, the little groups of
the mentioned points $U$, $S$, and $R$ comprise the whole space groups
$Pnma$ and $Pbam$, respectively.

Consequently, an antiferromagnetic structure with the magnetic groups
\begin{equation}
  \label{eq:12}
  M_{62} = Pnma + 
      \{K|0\textstyle\overline{\frac{1}{2}}\frac{1}{2}\}Pnma    
\end{equation}
and
\begin{equation}
  \label{eq:14}
  M_{55} = Pbam + 
      \{K|\textstyle\overline{\frac{1}{2}}\frac{1}{2}0\}Pbam    
\end{equation}
can be stable in BaFe$_2$As$_2$.

These two magnetic groups $M_{62}$~\gl{eq:12} and $M_{55}$~\gl{eq:14}
may be realized by the displacement of the Fe atoms as depicted in
Figs.~\ref{fig:structures} (c) and (d). By means of the generating
elements \gl{eq:9} and \gl{eq:11} we may verify that the symmetry
operations of the space groups $Pnma$ and $Pbam$ leave these
displacements of the Fe atoms invariant. In addition, the
displacements of the Fe atoms are also invariant under the
anti-unitary operations
$\{K|0\textstyle\overline{\frac{1}{2}}\frac{1}{2}\}$ and
$\{K|\textstyle\overline{\frac{1}{2}}\frac{1}{2}0\}$, respectively,
since they connect in any case two atoms with antiparallel magnetic
moments and the same displacement.

Consequently, the displacement of the Fe atoms depicted in
Figs.~\ref{fig:structures} (c) and (d) stabilize the experimentally
observed antiferromagnetic structure in BaFe$_2$As$_2$.

\section{Summary} 
\label{sec:result}
In Secs.~\ref{sec:64} and~\ref{sec:magneticband} of this paper we
could substantiate the existence of a magnetic band in the band
structure of BaFe$_2$As$_2$ which is related to the magnetic group
$M_{64}$~\gl{eq:3} of the experimentally observed~\cite{huang2}
antiferromagnetic state. As described by the nonadiabatic Heisenberg
model, the electrons in this extremely narrow and roughly
half-filled energy band can lower their Coulomb correlation energy by
producing this antiferromagnetic state.

However, this lowering of the energy is forbidden in undistorted
BaFe$_2$As$_2$ because a {\em stable} antiferromagnetic and its
time-inverted state need to form a basis of a suitable irreducible
corepresentation of the complete magnetic group of the Hamiltonian.
Such a corepresentation is not available in an {\em undistorted}
BaFe$_2$As$_2$ structure. However, as outlined in
Sec.~\ref{sec:subgroups}, a small shift of the Fe positions, brought
about by the electron system in the magnetic band, makes the required
corepresentation and thus a stable antiferromagnetic state.

We identified in Sec.~\ref{sec:stability} several distortions being
able to stabilize the antiferromagnetic state in BaFe$_2$As$_2$. The
two distortions with the highest symmetry are presented in
Figs.~\ref{fig:structures} (c) and (d).  They may be realized by the
depicted small displacements of the Fe atoms which are invariant under
the magnetic groups $M_{62}$~\gl{eq:12} and $M_{55}$~\gl{eq:14},
respectively. The related space groups $Pnma$ and $Pbam$ have the same
point group $D_{2h}$ as $Cmca$, but a lower translation symmetry: the
translation $\bm T_1$ in Fig.~\ref{fig:structures} (b) is no longer a
lattice translation in the distorted system, see
Figs.~\ref{fig:structures} (c) and (d). We suppose that these two
distortions of minor change of the symmetry are energetically more
favorable than any other possible distortion of lower symmetry.

\begin{acknowledgments} 
  We are indebted to Franz-Werner Gergen and Heinz Sch\"uhle from the
  EDV group of the Max-Planck-Institut f\"ur Intelligente Systeme and
  Kilian Krause from the TIK of the University in Stuttgart for their
  assistance in getting to run the computer programs needed for this
  work.
\end{acknowledgments} 

\setcounter{table}{0}
\renewcommand{\thetable}{A.\arabic{table}}


%

\appendix*
\section{Group-theoretical tables}
\label{sec:tables}
\onecolumngrid
\FloatBarrier
\begin{table*}[t]
\caption{Relationship between the lines of the Brillouin zones for the
  paramagnetic (Bravais lattice $\Gamma^v_q$) and the
  antiferromagnetic (Bravais lattice $\Gamma^b_o$)
  phases of undistorted BaFe$_2$As$_2$.  
\label{tab:falten}
} 
\begin{tabular}[t]{cp{.2cm}cp{.2cm}cp{.2cm}c}
\hline
$\Delta_M\Gamma$, $\Delta_M$X&&$\Gamma$Z, XPX && ZU$_M$, XU$_M$ && $\Gamma\Delta$X, XYUZ\\ 
 Z$\Gamma$ && $\Gamma$Y &&  YT && $\Gamma\Delta$FY\\
\hline
\end{tabular}
\ \\
\begin{flushleft}
Notes to Table~\ref{tab:falten}
\end{flushleft}
\begin{enumerate}
\item Always two lines in the paramagnetic Brillouin zone are
  equivalent to one line in the antiferromagnetic Brillouin zone.
\item The upper row lists pairs of lines in the Brillouin zone for the
  tetragonal paramagnetic phase. Beneath each pair the equivalent line in the
  Brillouin zone for the antiferromagnetic structure is listed.
\item The notations of the points and lines of symmetry follow
  Fig. 3.10. (b) (for $\Gamma^v_q$) and Fig. 3.6. (a) (for
  $\Gamma^b_o$) of Ref.~\cite{bc}.
\item XYUZ stands, e.g., for the line between X and Z via the lines Y
  and U in the Brillouin zone for $\Gamma^v_q$,
  cf. Fig.~\ref{fig:bandstr1}.
\item $\Delta_M$ and U$_M$ stand for the central points of the lines
  $\Gamma\Delta$X and XYUZ, respectively, in the Brillouin zone for
  $\Gamma^v_q$.
\end{enumerate}
\end{table*}

\FloatBarrier

\begin{table}[!]
\caption{
  Character tables of the single-valued irreducible representations of the
  space group $I4/mmm = \Gamma^v_qD^{17}_{4h}$ (139) of tetragonal
  paramagnetic BaFe$_2$As$_2$. 
\label{tab:rep139}}
\begin{tabular}[t]{ccccccccccc}
\multicolumn{11}{c}{$\Gamma (000)$\ and $Z (\frac{1}{2}\frac{1}{2}\overline{\frac{1}{2}})$}\\
 & $$ & $$ & $C^-_{4z}$ & $C_{2y}$ & $C_{2b}$ & $$ & $$ & $S^+_{4z}$ & $\sigma_y$ & $\sigma_{db}$\\
 & $E$ & $C_{2z}$ & $C^+_{4z}$ & $C_{2x}$ & $C_{2a}$ & $I$ & $\sigma_z$ & $S^-_{4z}$ & $\sigma_x$ & $\sigma_{da}$\\
\hline
$\Gamma^+_1$, $Z^+_1$ & 1 & 1 & 1 & 1 & 1 & 1 & 1 & 1 & 1 & 1\\
$\Gamma^+_2$, $Z^+_2$ & 1 & 1 & 1 & -1 & -1 & 1 & 1 & 1 & -1 & -1\\
$\Gamma^+_3$, $Z^+_3$ & 1 & 1 & -1 & 1 & -1 & 1 & 1 & -1 & 1 & -1\\
$\Gamma^+_4$, $Z^+_4$ & 1 & 1 & -1 & -1 & 1 & 1 & 1 & -1 & -1 & 1\\
$\Gamma^+_5$, $Z^+_5$ & 2 & -2 & 0 & 0 & 0 & 2 & -2 & 0 & 0 & 0\\
$\Gamma^-_1$, $Z^-_1$ & 1 & 1 & 1 & 1 & 1 & -1 & -1 & -1 & -1 & -1\\
$\Gamma^-_2$, $Z^-_2$ & 1 & 1 & 1 & -1 & -1 & -1 & -1 & -1 & 1 & 1\\
$\Gamma^-_3$, $Z^-_3$ & 1 & 1 & -1 & 1 & -1 & -1 & -1 & 1 & -1 & 1\\
$\Gamma^-_4$, $Z^-_4$ & 1 & 1 & -1 & -1 & 1 & -1 & -1 & 1 & 1 & -1\\
$\Gamma^-_5$, $Z^-_5$ & 2 & -2 & 0 & 0 & 0 & -2 & 2 & 0 & 0 & 0\\
\hline\\
\end{tabular}\hspace{1cm}
\begin{tabular}[t]{ccccccccc}
\multicolumn{9}{c}{$X (00\frac{1}{2})$}\\
 & $E$ & $C_{2z}$ & $C_{2a}$ & $C_{2b}$ & $I$ & $\sigma_z$ & $\sigma_{da}$ & $\sigma_{db}$\\
\hline
$X^+_1$ & 1 & 1 & 1 & 1 & 1 & 1 & 1 & 1\\
$X^+_2$ & 1 & -1 & 1 & -1 & 1 & -1 & 1 & -1\\
$X^+_3$ & 1 & 1 & -1 & -1 & 1 & 1 & -1 & -1\\
$X^+_4$ & 1 & -1 & -1 & 1 & 1 & -1 & -1 & 1\\
$X^-_1$ & 1 & 1 & 1 & 1 & -1 & -1 & -1 & -1\\
$X^-_2$ & 1 & -1 & 1 & -1 & -1 & 1 & -1 & 1\\
$X^-_3$ & 1 & 1 & -1 & -1 & -1 & -1 & 1 & 1\\
$X^-_4$ & 1 & -1 & -1 & 1 & -1 & 1 & 1 & -1\\
\hline\\
\end{tabular}\hspace{1cm}
\begin{tabular}[t]{cccccc}
\multicolumn{6}{c}{$P (\frac{1}{4}\frac{1}{4}\frac{1}{4})$}\\
 & $$ & $$ & $S^-_{4z}$ & $C_{2y}$ & $\sigma_{da}$\\
 & $E$ & $C_{2z}$ & $S^+_{4z}$ & $C_{2x}$ & $\sigma_{db}$\\
\hline
$P_1$ & 1 & 1 & 1 & 1 & 1\\
$P_2$ & 1 & 1 & 1 & -1 & -1\\
$P_3$ & 1 & 1 & -1 & 1 & -1\\
$P_4$ & 1 & 1 & -1 & -1 & 1\\
$P_5$ & 2 & -2 & 0 & 0 & 0\\
\hline\\
\end{tabular}\hspace{1cm}
\begin{tabular}[t]{ccccc}
\multicolumn{5}{c}{$N (0\frac{1}{2}0)$}\\
 & $E$ & $C_{2y}$ & $I$ & $\sigma_y$\\
\hline
$N^+_1$ & 1 & 1 & 1 & 1\\
$N^-_1$ & 1 & 1 & -1 & -1\\
$N^+_2$ & 1 & -1 & 1 & -1\\
$N^-_2$ & 1 & -1 & -1 & 1\\
\hline\\
\end{tabular}
\ \\
\begin{flushleft}
Notes to Table~\ref{tab:rep139}
\end{flushleft}
\begin{enumerate}
\item The space group $I4/mmm$ is symmorphic. Thus, the point group
  operations alone are symmetry operations.
\item The point group operations are related to the $x, y, z$
  coordinates in Fig.~\ref{fig:structures} (a).
\item The notations of the points of symmetry follow Fig. 3.10. (b) of 
Ref.~\cite{bc}.
\item The character table is determined from Table 5.7 of
  Ref.~\protect\cite{bc}. 
\end{enumerate}
\end{table}


\FloatBarrier
\begin{table}[!]
\caption{
Compatibility relations between points and lines in the Brillouin zone for
tetragonal paramagnetic BaFe$_2$As$_2$ (space group $I4/mmm$).  
\label{tab:komprel139}
}
\begin{tabular}[t]{cccccccccc}
\multicolumn{10}{c}{$\Gamma (000)$}\\
\hline
$\Gamma^+_1$ & $\Gamma^+_2$ & $\Gamma^+_3$ & $\Gamma^+_4$ & $\Gamma^+_5$ & $\Gamma^-_1$ & $\Gamma^-_2$ & $\Gamma^-_3$ & $\Gamma^-_4$ & $\Gamma^-_5$\\
$\Delta_1$ & $\Delta_2$ & $\Delta_2$ & $\Delta_1$ & $\Delta_3$ + $\Delta_4$ & $\Delta_3$ & $\Delta_4$ & $\Delta_4$ & $\Delta_3$ & $\Delta_1$ + $\Delta_2$\\
\hline\\
\end{tabular}\hspace{1cm}
\begin{tabular}[t]{cccccccc}
\multicolumn{8}{c}{$X (00\frac{1}{2})$}\\
\hline
$X^+_1$ & $X^+_2$ & $X^+_3$ & $X^+_4$ & $X^-_1$ & $X^-_2$ & $X^-_3$ & $X^-_4$\\
$\Delta_1$ & $\Delta_3$ & $\Delta_2$ & $\Delta_4$ & $\Delta_3$ & $\Delta_1$ & $\Delta_4$ & $\Delta_2$\\
\hline\\
\end{tabular}\hspace{1cm}
\begin{tabular}[t]{cccccccccc}
\multicolumn{10}{c}{$Z (\frac{1}{2}\frac{1}{2}\overline{\frac{1}{2}})$}\\
\hline
$Z^+_1$ & $Z^+_2$ & $Z^+_3$ & $Z^+_4$ & $Z^+_5$ & $Z^-_1$ & $Z^-_2$ & $Z^-_3$ & $Z^-_4$ & $Z^-_5$\\
$U_1$ & $U_2$ & $U_2$ & $U_1$ & $U_3$ + $U_4$ & $U_3$ & $U_4$ & $U_4$ & $U_3$ & $U_1$ + $U_2$\\
\hline\\
\end{tabular}\hspace{1cm}
\begin{tabular}[t]{cccccccc}
\multicolumn{8}{c}{$X' (\frac{1}{2}\frac{1}{2}0)$}\\
\hline
$X^+_1$ & $X^+_2$ & $X^+_3$ & $X^+_4$ & $X^-_1$ & $X^-_2$ & $X^-_3$ & $X^-_4$\\
$Y_1$ & $Y_4$ & $Y_2$ & $Y_3$ & $Y_3$ & $Y_2$ & $Y_4$ & $Y_1$\\
\hline\\
\end{tabular}\hspace{1cm}
\ \\
\begin{flushleft}
Notes to Table~\ref{tab:komprel139}
\end{flushleft}
\begin{enumerate}
\item The notations of the points and 
  lines of symmetry follow Fig. 3.10. (b) of Ref.~\cite{bc}.
\item $Z (\frac{1}{2}\frac{1}{2}\overline{\frac{1}{2}})$ and $X'
  (\frac{1}{2}\frac{1}{2}0)$ are connected  
via the lines $U$ and $Y$. 
The symmetry labels are chosen in such a way that $U_i = Y_i$ 
(for $i = 1,2,3,4$). 
\item 
This table defines the labels $\Delta_i$ and $U_i$ which are
needed in Table~\ref{tab:falten139_64} to find the representations of the
magnetic band at $Z$ and $T$.
\end{enumerate}
\end{table}


\FloatBarrier

\begin{table}[t]
\caption{
  Character tables of the single-valued irreducible representations of the
  orthorhombic space group $Cmca = \Gamma^b_oD^{18}_{2h}$ (64) of the
  experimentally observed~\cite{huang2} antiferromagnetic
  structure depicted in Fig.~\ref{fig:structures} (b).  
  \label{tab:rep64}}
\begin{tabular}[t]{ccccccccc}
\multicolumn{9}{c}{$\Gamma (000)$\ and $Y (\frac{1}{2}\frac{1}{2}0)$}\\
 & $\{E|000\}$ & $\{C_{2b}|\frac{1}{2}\frac{1}{2}\frac{1}{2}\}$ & $\{C_{2a}|\frac{1}{2}\frac{1}{2}\frac{1}{2}\}$ & $\{C_{2z}|000\}$ & $\{I|000\}$ & $\{\sigma_{db}|\frac{1}{2}\frac{1}{2}\frac{1}{2}\}$ & $\{\sigma_{da}|\frac{1}{2}\frac{1}{2}\frac{1}{2}\}$ & $\{\sigma_z|000\}$\\
\hline
$\Gamma^+_1$, $Y^+_1$ & 1 & 1 & 1 & 1 & 1 & 1 & 1 & 1\\
$\Gamma^+_2$, $Y^+_2$ & 1 & -1 & 1 & -1 & 1 & -1 & 1 & -1\\
$\Gamma^+_3$, $Y^+_3$ & 1 & 1 & -1 & -1 & 1 & 1 & -1 & -1\\
$\Gamma^+_4$, $Y^+_4$ & 1 & -1 & -1 & 1 & 1 & -1 & -1 & 1\\
$\Gamma^-_1$, $Y^-_1$ & 1 & 1 & 1 & 1 & -1 & -1 & -1 & -1\\
$\Gamma^-_2$, $Y^-_2$ & 1 & -1 & 1 & -1 & -1 & 1 & -1 & 1\\
$\Gamma^-_3$, $Y^-_3$ & 1 & 1 & -1 & -1 & -1 & -1 & 1 & 1\\
$\Gamma^-_4$, $Y^-_4$ & 1 & -1 & -1 & 1 & -1 & 1 & 1 & -1\\
\hline\\
\end{tabular}\hspace{1cm}
\begin{tabular}[t]{ccccccccccc}
\multicolumn{11}{c}{$Z (00\frac{1}{2})$\ and $T (\frac{1}{2}\frac{1}{2}\frac{1}{2})$}\\
 & $$ & $$ & $\{C_{2b}|\frac{1}{2}\frac{1}{2}\frac{3}{2}\}$ & $\{I|001\}$ & $\{\sigma_{db}|\frac{1}{2}\frac{1}{2}\frac{3}{2}\}$ & $$ & $$ & $\{\sigma_{da}|\frac{1}{2}\frac{1}{2}\frac{3}{2}\}$ & $\{C_{2z}|001\}$ & $\{C_{2a}|\frac{1}{2}\frac{1}{2}\frac{3}{2}\}$\\
 & $\{E|000\}$ & $\{E|001\}$ & $\{C_{2b}|\frac{1}{2}\frac{1}{2}\frac{1}{2}\}$ & $\{I|000\}$ & $\{\sigma_{db}|\frac{1}{2}\frac{1}{2}\frac{1}{2}\}$ & $\{\sigma_z|000\}$ & $\{\sigma_z|001\}$ & $\{\sigma_{da}|\frac{1}{2}\frac{1}{2}\frac{1}{2}\}$ & $\{C_{2z}|000\}$ & $\{C_{2a}|\frac{1}{2}\frac{1}{2}\frac{1}{2}\}$\\
\hline
$Z_1$, $T_1$ & 2 & -2 & 0 & 0 & 0 & 2 & -2 & 0 & 0 & 0\\
$Z_2$, $T_2$ & 2 & -2 & 0 & 0 & 0 & -2 & 2 & 0 & 0 & 0\\
\hline\\
\end{tabular}\hspace{1cm}
\begin{tabular}[t]{ccccccccccc}
\multicolumn{11}{c}{$R (0\frac{1}{2}\frac{1}{2})$}\\
 & $K$ & $\{K|\frac{1}{2}\overline{\frac{1}{2}}\frac{1}{2}\}$ & $\{E|000\}$ & $\{\sigma_{db}|\frac{1}{2}\frac{1}{2}\frac{1}{2}\}$ & $\{E|001\}$ & $\{\sigma_{db}|\frac{1}{2}\frac{1}{2}\frac{3}{2}\}$ & $\{I|001\}$ & $\{C_{2b}|\frac{1}{2}\frac{1}{2}\frac{3}{2}\}$ & $\{I|000\}$ & $\{C_{2b}|\frac{1}{2}\frac{1}{2}\frac{1}{2}\}$\\
\hline
$R^+_1$ & (c) & (a) & 1 & i & -1 & -i & 1 & i & -1 & -i\\
$R^+_2$ & (c) & (a) & 1 & -i & -1 & i & 1 & -i & -1 & i\\
$R^-_1$ & (c) & (a) & 1 & i & -1 & -i & -1 & -i & 1 & i\\
$R^-_2$ & (c) & (a) & 1 & -i & -1 & i & -1 & i & 1 & -i\\
\hline\\
\end{tabular}\hspace{1cm}

\hspace{1cm}
\ \\
\begin{flushleft}
Notes to Table~\ref{tab:rep64}
\end{flushleft}
\begin{enumerate}
\item The notations of the points of symmetry follow Fig. 3.6. (a) 
of Ref.~\cite{bc}.
\item Point $S$ is not listed because it has only one two-dimensional
  representation $S_1$ irrelevant in this paper. 
\item The space group operations are related to the coordinate system in
  Fig.~\ref{fig:structures} (b).
\item The character tables are determined from Table 5.7 in
  Ref.~\protect\cite{bc}. The origin of the coordinate system used in
  Ref.~\protect\cite{bc} for $Cmca$ is translated into the origin used
  in this paper by $\bm t_0 = \frac{1}{4}\bm T_1 + \frac{1}{4}\bm
  T_2$. Thus, the symmetry operations $P_{bc}$ given in Table 5.7 of
  Ref.~\protect\cite{bc} are changed into the operations $P_{p}$ used
  in this paper by the equation
  $$P_{p} = \{E|\bm t_0\}P_{bc}\{E|-\bm t_0\}, $$
  where $E$ is the indentity operation. 
\item The cases (a) and (c) are defined in equations\ (7.3.45) and
  (7.3.47), respectively, of Ref.~\cite{bc} and are determined by
  equation\ (7.3.51) of Ref.~\cite{bc}.
\end{enumerate}
\end{table}


\FloatBarrier

\begin{table}[!]
\caption{
Compatibility relations between the Brillouin zone for tetragonal
paramagnetic BaFe$_2$As$_2$ and the Brillouin zone for the orthorhombic
antiferromagnetic structure depicted in Figs.~\ref{fig:structures} (a) and
(b), respectively.  
\label{tab:falten139_64}
}
\begin{tabular}[t]{cccccccccc}
\multicolumn{10}{c}{$\Gamma (000)$}\\
\hline
$\Gamma^+_1$ & $\Gamma^+_2$ & $\Gamma^+_3$ & $\Gamma^+_4$ & $\Gamma^+_5$ & $\Gamma^-_1$ & $\Gamma^-_2$ & $\Gamma^-_3$ & $\Gamma^-_4$ & $\Gamma^-_5$\\
$\Gamma^+_1$ & $\Gamma^+_4$ & $\Gamma^+_4$ & $\Gamma^+_1$ & $\Gamma^+_2$ + $\Gamma^+_3$ & $\Gamma^-_1$ & $\Gamma^-_4$ & $\Gamma^-_4$ & $\Gamma^-_1$ & $\Gamma^-_2$ + $\Gamma^-_3$\\
\hline\\
\end{tabular}\hspace{1cm}
\begin{tabular}[t]{cccccccc}
\multicolumn{8}{c}{$X (00\frac{1}{2})$}\\
\hline
$X^+_1$ & $X^+_2$ & $X^+_3$ & $X^+_4$ & $X^-_1$ & $X^-_2$ & $X^-_3$ & $X^-_4$\\
$Y^+_1$ & $Y^+_2$ & $Y^+_4$ & $Y^+_3$ & $Y^-_1$ & $Y^-_2$ & $Y^-_4$ & $Y^-_3$\\
\hline\\
\end{tabular}\hspace{1cm}
\begin{tabular}[t]{cccccccccc}
\multicolumn{10}{c}{$Z (\frac{1}{2}\frac{1}{2}\overline{\frac{1}{2}})$}\\
\hline
$Z^+_1$ & $Z^+_2$ & $Z^+_3$ & $Z^+_4$ & $Z^+_5$ & $Z^-_1$ & $Z^-_2$ & $Z^-_3$ & $Z^-_4$ & $Z^-_5$\\
$Y^+_4$ & $Y^+_1$ & $Y^+_1$ & $Y^+_4$ & $Y^+_2$ + $Y^+_3$ & $Y^-_4$ & $Y^-_1$ & $Y^-_1$ & $Y^-_4$ & $Y^-_2$ + $Y^-_3$\\
\hline\\
\end{tabular}\hspace{1cm}
\begin{tabular}[t]{cccccccc}
\multicolumn{8}{c}{$X' (\frac{1}{2}\overline{\frac{1}{2}}0)$}\\
\hline
$X^+_1$ & $X^+_2$ & $X^+_3$ & $X^+_4$ & $X^-_1$ & $X^-_2$ & $X^-_3$ & $X^-_4$\\
$\Gamma^+_4$ & $\Gamma^+_2$ & $\Gamma^+_1$ & $\Gamma^+_3$ & $\Gamma^-_4$ & $\Gamma^-_2$ & $\Gamma^-_1$ & $\Gamma^-_3$\\
\hline\\
\end{tabular}\hspace{1cm}
\begin{tabular}[t]{cccc}
\multicolumn{4}{c}{$N' (\frac{1}{2}0\overline{\frac{1}{2}})$}\\
\hline
$N^+_1$ & $N^-_1$ & $N^+_2$ & $N^-_2$\\
$R^-_1$ + $R^-_2$ & $R^+_1$ + $R^+_2$ & $R^-_1$ + $R^-_2$ & $R^+_1$ + $R^+_2$\\
\hline\\
\end{tabular}\hspace{1cm}
\begin{tabular}[t]{cccc}
\multicolumn{4}{c}{$\Delta_M' (\frac{1}{4}\overline{\frac{1}{4}}0)$}\\
\hline
$\Delta_1$ & $\Delta_3$ & $\Delta_2$ & $\Delta_4$\\
$Z_1$ & $Z_2$ & $Z_1$ & $Z_2$\\
\hline\\
\end{tabular}\hspace{1cm}
\begin{tabular}[t]{cccc}
\multicolumn{4}{c}{$U_M' (\frac{3}{4}\frac{1}{4}\overline{\frac{1}{2}})$}\\
\hline
$U_1$ & $U_3$ & $U_2$ & $U_4$\\
$T_1$ & $T_2$ & $T_1$ & $T_2$\\
\hline\\
\end{tabular}\hspace{1cm}
\ \\
\begin{flushleft}
Notes to Table~\ref{tab:falten139_64}
\end{flushleft}
\begin{enumerate}
\item The antiferromagnetic structure in Fig.~\ref{fig:structures} (b) has the
  space group $Cmca$.
\item The Brillouin zone for the orthorhombic space group $Cmca$ lies
  within the Brillouin zone for the tetragonal space group $I4/mmm$.
\item $\Delta_M(00\frac{1}{4})$ and 
$U_M(\frac{1}{2}\frac{1}{2}\overline{\frac{1}{4}})$ stand for the central
points of the lines connecting $\Gamma$ with $X(00\frac{1}{2})$ and
$Z(\frac{1}{2}\frac{1}{2}\overline{\frac{1}{2}})$ with
$X'(\frac{1}{2}\frac{1}{2}0)$, respectively.
\item The points $X'$, $N'$, $\Delta'_M$, and $U'_M$ do not belong to the
  basic domain of the Brillouin zone but can be generated by the action of 
  the point group operations $C^+_{4z}$, $C_{2z}$, $C^+_{4z}$, and $C^+_{4z}$
  on $X$, $N$, $\Delta_M$, and $U_M$, respectively.
\item The upper rows list the representations of the little groups of the
  points of symmetry in the Brillouin zone for the tetragonal paramagnetic
  phase. The lower rows list
  representations of the little groups of the related points of symmetry in
  the Brillouin zone for the antiferromagnetic structure.
  
  The representations in the same column are compatible in the
  following sense: Bloch functions that are basis functions of a
  representation $D_i$ in the upper row can be unitarily transformed into
  the basis functions of the representation given below $D_i$.
\item The symmetry labels for the points of symmetry are given in
  Tables~\ref{tab:rep139} and~\ref{tab:rep64}, respectively.
\item The symmetry labels for the lines $\Delta$ and
  $U$ are defined by Table~\ref{tab:komprel139}.  
\item The compatibility relations are determined in the way described in
  great detail in Ref.~\cite{eabf}.
\end{enumerate}
\end{table}


\FloatBarrier
\begin{table}[!]
\caption{
Compatibility relations between the points $\Gamma (000)$ and 
$Y (\overline{\frac{1}{2}}\frac{1}{2}0)$ and the line connecting these points
in the Brillouin zone for the antiferromagnetic structure in undistorted 
BaFe$_2$As$_2$ (space group $Cmca$).  
\label{tab:komprel64}
}
\begin{tabular}[t]{cccccccc}
\multicolumn{8}{c}{$\Gamma (000)$}\\
\hline
$\Gamma^+_1$ & $\Gamma^+_2$ & $\Gamma^+_3$ & $\Gamma^+_4$ & $\Gamma^-_1$ & $\Gamma^-_2$ & $\Gamma^-_3$ & $\Gamma^-_4$\\
$\Delta_1$ & $\Delta_3$ & $\Delta_4$ & $\Delta_2$ & $\Delta_3$ & $\Delta_1$ & $\Delta_2$ & $\Delta_4$\\
\hline\\
\end{tabular}\hspace{1cm}
\begin{tabular}[t]{cccccccc}
\multicolumn{8}{c}{$Y (\overline{\frac{1}{2}}\frac{1}{2}0)$}\\
\hline
$Y^+_1$ & $Y^+_2$ & $Y^+_3$ & $Y^+_4$ & $Y^-_1$ & $Y^-_2$ & $Y^-_3$ & $Y^-_4$\\
$F_1$ & $F_3$ & $F_4$ & $F_2$ & $F_3$ & $F_1$ & $F_2$ & $F_4$\\
\hline\\
\end{tabular}\hspace{1cm}
\ \\
\begin{flushleft}
Notes to Table~\ref{tab:komprel64}
\end{flushleft}
\begin{enumerate}
\item The notations of the points and 
  lines of symmetry follow Fig. 3.6. (a) of Ref.~\cite{bc}.
\item $\Gamma (000)$ and $Y (\overline{\frac{1}{2}}\frac{1}{2}0)$ are connected 
via the lines $\Delta$ and $F$. 
The symmetry notations are chosen in such a way that $\Delta_i = F_i$ 
(for $i = 1,2,3,4$). 
\item 
This table defines the labels $\Delta_i$ and $F_i$ which are needed to 
construct the magnetic band between $\Gamma (000)$ and 
$Y (\overline{\frac{1}{2}}\frac{1}{2}0)$ in Fig.~\ref{fig:bandstr2}. 
\end{enumerate}
\end{table}


\FloatBarrier

\begin{table}
\caption{
Single-valued representations of all the energy bands of antiferromagnetic 
BaFe$_2$As$_2$ with symmetry-adapted and optimally  
localized Wannier functions centered at the Fe atoms. 
\label{tab:wf64}}
\begin{tabular}[t]{ccccccc}
Fe & $\Gamma$ & $Y$ & $Z$ & $T$ & $R$ \\
\hline
Band 1 & $\Gamma^+_1$ + $\Gamma^+_2$ + $\Gamma^-_1$ + $\Gamma^-_2$ & $Y^+_1$ + $Y^+_2$ + $Y^-_1$ + $Y^-_2$ & $Z_1$ + $Z_2$ & $T_1$ + $T_2$ & $R^+_1$ + $R^+_2$ + $R^-_1$ + $R^-_2$ \\
Band 2 & $\Gamma^+_3$ + $\Gamma^+_4$ + $\Gamma^-_3$ + $\Gamma^-_4$ & $Y^+_3$ + $Y^+_4$ + $Y^-_3$ + $Y^-_4$ & $Z_1$ + $Z_2$ & $T_1$ + $T_2$ & $R^+_1$ + $R^+_2$ + $R^-_1$ + $R^-_2$ \\
\hline\\
\end{tabular}\hspace{1cm}
\ \\
\begin{flushleft}
Notes to Table~\ref{tab:wf64}
\end{flushleft}
\begin{enumerate}
\item The antiferromagnetic structure of undistorted BaFe$_2$As$_2$ depicted in 
Fig.~\ref{fig:structures} (b) has the space group $Cmca$ and the magnetic
group $M = Cmca + \{K|\frac{1}{2}\overline{\frac{1}{2}}\frac{1}{2}\}Cmca$ with
$K$ denoting the operator of time-inversion.
\item The two listed bands form ``magnetic bands'' related to $M$.   
\item Each row defines one band consisting of four branches, because there
  are four Fe atoms in the unit cell.
\item The representations are given in Table~\ref{tab:rep64}.
\item The bands are determined by Eq.~(23) of Ref.~\protect\cite{josla2cuo4}.
\item Assume the tetragonal band structure of BaFe$_2$As$_2$ to be
  folded into the Brillouin zone for the antiferromagnetic phase (as carried
  out in Fig.~\ref{fig:bandstr2}) and assume further a band of the symmetry in any row of this table to
  exist in the folded band structure. Then the Bloch functions of this  
  band can be unitarily transformed into Wannier functions that are
\begin{itemize}
\item as well localized as possible; 
\item centered at the Fe atoms;
\item and symmetry-adapted to the space group $Cmca$ of the antiferromagnetic phase.
\end{itemize}
\item Eq.~(23) of Ref.~\protect\cite{josla2cuo4} makes sure that the Wannier
  function may be chosen to be symmetry-adapted to the {\em space} group $Cmca$ of the antiferromagnetic phase. In
  addition, there exists a Matrix {\bf N}, namely  
$ {\bf N} = 
  {\tiny
  \left(
  \begin{array}{cccc} 
  0 & 0 & 1 & 0\\              
  0 & 0 & 0 & 1\\              
  1 & 0 & 0 & 0\\              
  0 & 1 & 0 & 0\\              
  \end{array}
  \right)
  } $,
 satisfying both Eqs.~(26) (with
  $\{K|\frac{1}{2}\overline{\frac{1}{2}}\frac{1}{2}\}$) and~(32) of Ref.~\protect\cite{josla2cuo4}
  for the two bands listed in this table. Hence, the Wannier functions may be
  chosen symmetry adapted to the magnetic group $M = Cmca +
  \{K|\frac{1}{2}\overline{\frac{1}{2}}\frac{1}{2}\}Cmca$.
\end{enumerate}
\end{table}


\FloatBarrier

\begin{table}[t]
\caption{
  Character tables of the single-valued irreducible representations of the
  orthorhombic space group $Pnma = \Gamma_oD^{16}_{2h}$ (62).
  \label{tab:rep62}}
\begin{tabular}[t]{cccccccccc}
\multicolumn{10}{c}{$U (0\frac{1}{2}\frac{1}{2})$}\\
 & $K$ & $\{K|0\overline{\frac{1}{2}}\frac{1}{2}\}$ & $\{E|000\}$ & $\{C_{2a}|0\frac{1}{2}\frac{1}{2}\}$ & $\{E|001\}$ & $\{C_{2a}|0\frac{1}{2}\frac{3}{2}\}$ & $\{C_{2z}|\frac{1}{2}\frac{1}{2}0\}$ & $\{C_{2b}|\frac{1}{2}0\frac{3}{2}\}$ & $\{C_{2z}|\frac{1}{2}\frac{1}{2}1\}$\\
\hline
$U^-_1$ & (c) & (a) & 1 & i & -1 & -i & 1 & i & -1\\
$U^-_2$ & (c) & (a) & 1 & -i & -1 & i & 1 & -i & -1\\
$U^-_3$ & (c) & (a) & 1 & i & -1 & -i & -1 & -i & 1\\
$U^-_4$ & (c) & (a) & 1 & -i & -1 & i & -1 & i & 1\\
$U^+_1$ & (c) & (a) & 1 & i & -1 & -i & 1 & i & -1\\
$U^+_2$ & (c) & (a) & 1 & -i & -1 & i & 1 & -i & -1\\
$U^+_3$ & (c) & (a) & 1 & i & -1 & -i & -1 & -i & 1\\
$U^+_4$ & (c) & (a) & 1 & -i & -1 & i & -1 & i & 1\\
\hline\\
\end{tabular}\hspace{1cm}
\begin{tabular}[t]{cccccccccc}
\multicolumn{10}{c}{$U (0\frac{1}{2}\frac{1}{2})$\qquad $(continued)$}\\
 & $\{C_{2b}|\frac{1}{2}0\frac{1}{2}\}$ & $\{I|\frac{1}{2}\frac{1}{2}0\}$ & $\{\sigma_{da}|\frac{1}{2}0\frac{3}{2}\}$ & $\{I|\frac{1}{2}\frac{1}{2}1\}$ & $\{\sigma_{da}|\frac{1}{2}0\frac{1}{2}\}$ & $\{\sigma_z|000\}$ & $\{\sigma_{db}|0\frac{1}{2}\frac{1}{2}\}$ & $\{\sigma_z|001\}$ & $\{\sigma_{db}|0\frac{1}{2}\frac{3}{2}\}$\\
\hline
$U^-_1$ & -i & 1 & i & -1 & -i & 1 & i & -1 & -i\\
$U^-_2$ & i & 1 & -i & -1 & i & 1 & -i & -1 & i\\
$U^-_3$ & i & 1 & i & -1 & -i & -1 & -i & 1 & i\\
$U^-_4$ & -i & 1 & -i & -1 & i & -1 & i & 1 & -i\\
$U^+_1$ & -i & -1 & -i & 1 & i & -1 & -i & 1 & i\\
$U^+_2$ & i & -1 & i & 1 & -i & -1 & i & 1 & -i\\
$U^+_3$ & i & -1 & -i & 1 & i & 1 & i & -1 & -i\\
$U^+_4$ & -i & -1 & i & 1 & -i & 1 & -i & -1 & i\\
\hline\\
\end{tabular}\hspace{1cm}

\hspace{1cm}
\ \\
\begin{flushleft}
Notes to Table~\ref{tab:rep62}
\end{flushleft}
\begin{enumerate}
\item The notations of the points of symmetry follow Fig. 3.5. of
  Ref.~\cite{bc}.
\item Only the listed point $U$ is relevant in this paper. 
\item The notation of the space group operations is related to the
  coordinate system in Fig.~\ref{fig:structures} (c).
\item The character tables are determined from Table 5.7 in
  Ref.~\protect\cite{bc}. The origin of the coordinate system used in
  Ref.~\protect\cite{bc} for $Pnma$ is identical with the origin used
  in this paper.
\item The cases (a) and (c) are defined in equations\ (7.3.45) and
  (7.3.47), respectively, of Ref.~\cite{bc} and are determined by
  equation\ (7.3.51) of Ref.~\cite{bc}
\end{enumerate}
\end{table}


\FloatBarrier

\begin{table}[t]
\caption{
  Character tables of the single-valued irreducible representations of the
  orthorhombic space group $Pbam = \Gamma_oD^{9}_{2h}$ (55).
  \label{tab:rep55}}
\begin{tabular}[t]{cccccccccc}
\multicolumn{10}{c}{$S (\overline{\frac{1}{2}}\frac{1}{2}0)$}\\
 & $K$ & $\{K|\overline{\frac{1}{2}}\frac{1}{2}0\}$ & $\{E|000\}$ & $\{C_{2b}|\frac{1}{2}\frac{1}{2}0\}$ & $\{E|010\}$ & $\{C_{2b}|\frac{1}{2}\frac{3}{2}0\}$ & $\{C_{2z}|000\}$ & $\{C_{2a}|\frac{1}{2}\frac{1}{2}0\}$ & $\{C_{2z}|010\}$\\
\hline
$S^-_1$ & (c) & (a) & 1 & i & -1 & -i & 1 & i & -1\\
$S^-_2$ & (c) & (a) & 1 & -i & -1 & i & 1 & -i & -1\\
$S^-_3$ & (c) & (a) & 1 & i & -1 & -i & -1 & -i & 1\\
$S^-_4$ & (c) & (a) & 1 & -i & -1 & i & -1 & i & 1\\
$S^+_1$ & (c) & (a) & 1 & i & -1 & -i & 1 & i & -1\\
$S^+_2$ & (c) & (a) & 1 & -i & -1 & i & 1 & -i & -1\\
$S^+_3$ & (c) & (a) & 1 & i & -1 & -i & -1 & -i & 1\\
$S^+_4$ & (c) & (a) & 1 & -i & -1 & i & -1 & i & 1\\
\hline\\
\end{tabular}\hspace{1cm}
\begin{tabular}[t]{cccccccccc}
\multicolumn{10}{c}{$S (\overline{\frac{1}{2}}\frac{1}{2}0)$\qquad $(continued)$}\\
 & $\{C_{2a}|\frac{1}{2}\frac{3}{2}0\}$ & $\{I|000\}$ & $\{\sigma_{db}|\frac{1}{2}\frac{1}{2}0\}$ & $\{I|010\}$ & $\{\sigma_{db}|\frac{1}{2}\frac{3}{2}0\}$ & $\{\sigma_z|000\}$ & $\{\sigma_{da}|\frac{1}{2}\frac{1}{2}0\}$ & $\{\sigma_z|010\}$ & $\{\sigma_{da}|\frac{1}{2}\frac{3}{2}0\}$\\
\hline
$S^-_1$ & -i & 1 & i & -1 & -i & 1 & i & -1 & -i\\
$S^-_2$ & i & 1 & -i & -1 & i & 1 & -i & -1 & i\\
$S^-_3$ & i & 1 & i & -1 & -i & -1 & -i & 1 & i\\
$S^-_4$ & -i & 1 & -i & -1 & i & -1 & i & 1 & -i\\
$S^+_1$ & -i & -1 & -i & 1 & i & -1 & -i & 1 & i\\
$S^+_2$ & i & -1 & i & 1 & -i & -1 & i & 1 & -i\\
$S^+_3$ & i & -1 & -i & 1 & i & 1 & i & -1 & -i\\
$S^+_4$ & -i & -1 & i & 1 & -i & 1 & -i & -1 & i\\
\hline\\
\end{tabular}\hspace{1cm}
\begin{tabular}[t]{cccccccccc}
\multicolumn{10}{c}{$R (\overline{\frac{1}{2}}\frac{1}{2}\frac{1}{2})$}\\
 & $K$ & $\{K|\overline{\frac{1}{2}}\frac{1}{2}0\}$ & $\{E|000\}$ & $\{C_{2b}|\frac{1}{2}\frac{1}{2}0\}$ & $\{E|001\}$ & $\{C_{2b}|\frac{1}{2}\frac{1}{2}1\}$ & $\{C_{2z}|000\}$ & $\{C_{2a}|\frac{1}{2}\frac{1}{2}0\}$ & $\{C_{2z}|001\}$\\
\hline
$R^-_1$ & (c) & (a) & 1 & i & -1 & -i & 1 & i & -1\\
$R^-_2$ & (c) & (a) & 1 & -i & -1 & i & 1 & -i & -1\\
$R^-_3$ & (c) & (a) & 1 & i & -1 & -i & -1 & -i & 1\\
$R^-_4$ & (c) & (a) & 1 & -i & -1 & i & -1 & i & 1\\
$R^+_1$ & (c) & (a) & 1 & i & -1 & -i & 1 & i & -1\\
$R^+_2$ & (c) & (a) & 1 & -i & -1 & i & 1 & -i & -1\\
$R^+_3$ & (c) & (a) & 1 & i & -1 & -i & -1 & -i & 1\\
$R^+_4$ & (c) & (a) & 1 & -i & -1 & i & -1 & i & 1\\
\hline\\
\end{tabular}\hspace{1cm}
\begin{tabular}[t]{cccccccccc}
\multicolumn{10}{c}{$R (\overline{\frac{1}{2}}\frac{1}{2}\frac{1}{2})$\qquad $(continued)$}\\
 & $\{C_{2a}|\frac{1}{2}\frac{1}{2}1\}$ & $\{I|000\}$ & $\{\sigma_{db}|\frac{1}{2}\frac{1}{2}0\}$ & $\{I|001\}$ & $\{\sigma_{db}|\frac{1}{2}\frac{1}{2}1\}$ & $\{\sigma_z|000\}$ & $\{\sigma_{da}|\frac{1}{2}\frac{1}{2}0\}$ & $\{\sigma_z|001\}$ & $\{\sigma_{da}|\frac{1}{2}\frac{1}{2}1\}$\\
\hline
$R^-_1$ & -i & 1 & i & -1 & -i & 1 & i & -1 & -i\\
$R^-_2$ & i & 1 & -i & -1 & i & 1 & -i & -1 & i\\
$R^-_3$ & i & 1 & i & -1 & -i & -1 & -i & 1 & i\\
$R^-_4$ & -i & 1 & -i & -1 & i & -1 & i & 1 & -i\\
$R^+_1$ & -i & -1 & -i & 1 & i & -1 & -i & 1 & i\\
$R^+_2$ & i & -1 & i & 1 & -i & -1 & i & 1 & -i\\
$R^+_3$ & i & -1 & -i & 1 & i & 1 & i & -1 & -i\\
$R^+_4$ & -i & -1 & i & 1 & -i & 1 & -i & -1 & i\\
\hline\\
\end{tabular}\hspace{1cm}
\hspace{1cm}
\ \\
\begin{flushleft}
Notes to Table~\ref{tab:rep55}
\end{flushleft}
\begin{enumerate}
\item The notations of the points of symmetry follow Fig. 3.5. of
  Ref.~\cite{bc}.
\item Only the listed points $S$ and $R$ are relevant in this paper. 
\item The notation of the space group operations is related to the
  coordinate system in Fig.~\ref{fig:structures} (d).
\item The character tables are determined from Table 5.7 in
  Ref.~\protect\cite{bc}. The origin of the coordinate system used in
  Ref.~\protect\cite{bc} for $Pbam$ is identical with the origin used
  in this paper.
\item The cases (a) and (c) are defined in equations\ (7.3.45) and
  (7.3.47), respectively, of Ref.~\cite{bc} and are determined by
  equation\ (7.3.51) of Ref.~\cite{bc}.
\end{enumerate}
\end{table}


\end{document}